\documentclass[aps,pre,twocolumn,floatfix,superscriptaddress]{revtex4-1}

\usepackage{amsmath}  
\usepackage{amsfonts} 
\usepackage{graphicx} 
\usepackage{epstopdf}
\usepackage{gensymb}
\usepackage{physics}
\usepackage{xcolor}
\DeclareMathOperator{\area}{\text{area}}

\newcommand{\markercirc}{\raisebox{-0.15\height}{\includegraphics{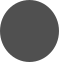}}}
\newcommand{\markerstar}{\raisebox{-0.15\height}{\includegraphics{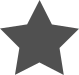}}}
\newcommand{\markersq}{\raisebox{-0.15\height}{\includegraphics{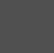}}}

\begin{document}
\title{Odd elasticity}

\author{Colin Scheibner}
\thanks{These authors contributed equally to this work.}
\affiliation{James Franck Institute, The University of Chicago, Chicago, Illinois 60637, USA}
\affiliation{Department of Physics, The University of Chicago, Chicago, Illinois 60637, USA}
\author{Anton Souslov}
\thanks{These authors contributed equally to this work.}
\affiliation{James Franck Institute, The University of Chicago, Chicago, Illinois 60637, USA}
\affiliation{Department of Physics, University of Bath, Bath BA2 7AY, United Kingdom}
\author{Debarghya Banerjee}
\affiliation{Max Planck Institute for Dynamics and Self-Organization, 37077 G\"{o}ttingen, Germany}
\affiliation{Instituut-Lorentz, Universiteit Leiden, Leiden 2300 RA, The Netherlands}
\author{Piotr Surowka}
\affiliation{Max Planck Institute for the Physics of Complex Systems, 01187 Dresden, Germany}
\author{William T.~M.~Irvine}
\affiliation{James Franck Institute, The University of Chicago, Chicago, Illinois 60637, USA}
\affiliation{Department of Physics, The University of Chicago, Chicago, Illinois 60637, USA}
\affiliation{Enrico Fermi Institute, The University of Chicago, Chicago, Illinois, 60637, USA}
\author{Vincenzo Vitelli}
\email{vitelli@uchicago.edu}
\affiliation{James Franck Institute, The University of Chicago, Chicago, Illinois 60637, USA}
\affiliation{Department of Physics, The University of Chicago, Chicago, Illinois 60637, USA}

\begin{abstract}
Hooke's law states that the forces or stresses experienced by an elastic object are proportional to the applied deformations or strains. The number of coefficients of proportionality between stress and strain, i.e., the elastic moduli, is constrained by energy conservation. In this Letter, we lift this restriction and generalize linear elasticity to active media with non-conservative microscopic interactions that violate mechanical reciprocity. 
This generalized framework, which we dub odd elasticity, reveals that two additional moduli can exist in a two-dimensional isotropic solid with active bonds. Such an odd-elastic solid can be regarded as a distributed engine: work is locally extracted, or injected, during quasi-static cycles of deformation. Using continuum equations, coarse-grained microscopic models, and numerical simulations, we uncover phenomena ranging from activity-induced auxetic behavior to wave propagation powered by self-sustained active elastic cycles. Besides providing insights beyond existing hydrodynamic theories of active solids, odd elasticity suggests design principles for emergent autonomous materials.
\end{abstract}

\maketitle

One of the central assumptions of classical elasticity is that the work needed to deform a material element depends only on its initial and final states~\cite{Landau7}. 
If the work is path dependent, the local stresses cannot be obtained from derivatives of an elastic potential energy. 
Nonetheless, even without an elastic potential, the stress-strain relation exists and can be linearized for small deformations. This approximation, known as Hooke's law, is valid for solids both in and out of equilibrium and can be captured by the equation $\sigma_{ij}=K_{ijmn}u_{mn}$, where $u_{mn}$ are the gradients $\partial_m u_n$ of the displacement vector $u_n$ and $K_{ijmn}$ is the stiffness tensor~\cite{Landau7}. 
In the absence of an elastic potential energy, we find that the most general linear stress-strain relation for an isotropic two-dimensional solid reads (see Methods):
\begin{equation}
    \raisebox{-0.5\height}{\includegraphics{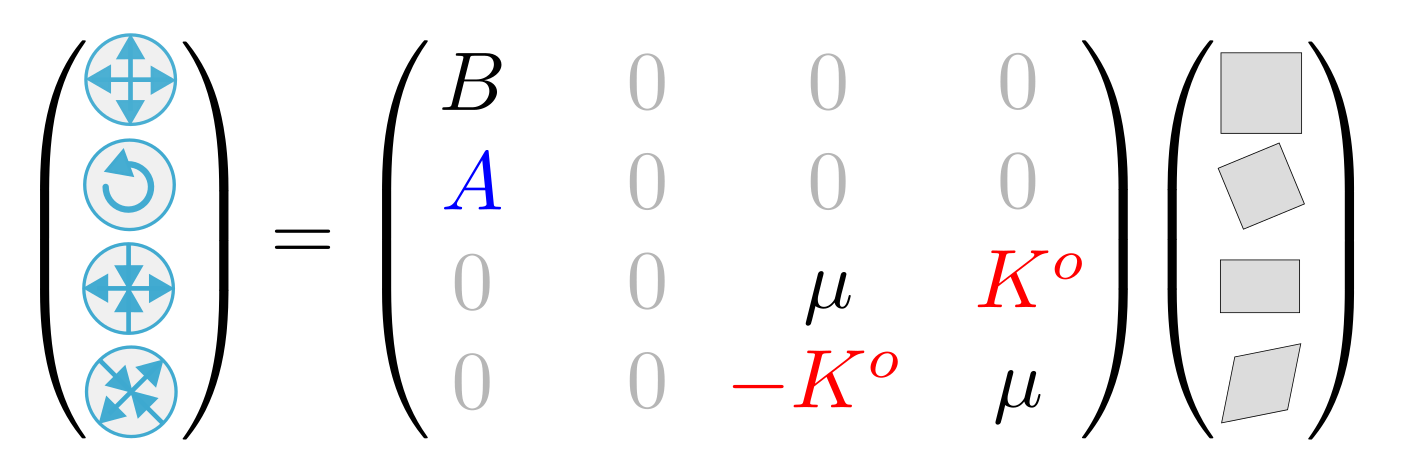}}
\end{equation}
The geometric notation in Eq.~(1) is illustrated in \mbox{Fig.~1a-b}. The displacement gradients on the right-hand side are decomposed along the four independent components shown in Fig.~1: dilation, rotation, and the two shear deformations $S_1$ and $S_2$. Similarly, the stress vector on the left-hand side is decomposed into pressure, torque, and the two shear stresses. 
The stiffness matrix represents the tensor $K_{ijmn}$ and contains all of the allowed elastic moduli~\cite{Landau7}. The entries in the second column are zero because no stress is generated by reorienting the solid. Besides the familiar bulk modulus, $B$, and shear modulus, $\mu$, there are two additional entries in Eq.~(1): $A$ and $K^o$. Qualitatively, the modulus $A$ couples compression (and dilation) to an internal torque density in the solid. By contrast, $K^o$ describes a nonreciprocal response in shear stress along a direction rotated with respect to the applied shear strain, but $K^o$ does not entail a net torque density in the solid, see Fig.~1 (a precise classification is provided in the Methods). 

 \begin{figure*}[h!]
 \includegraphics{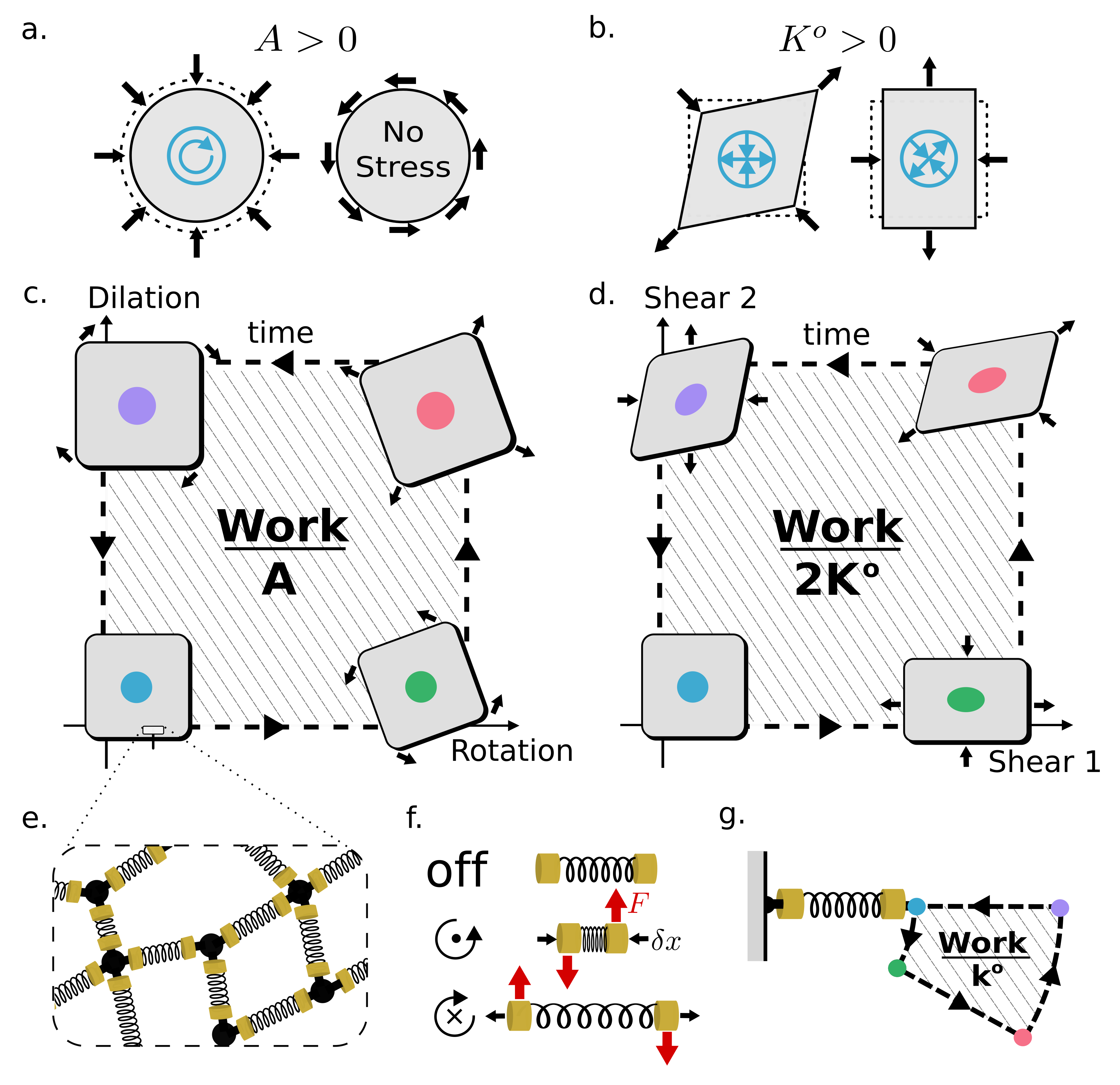}
 \caption{{\bf Active elastic engine cycle.} {\bf a.~} 
 If a material has a nonzero odd modulus $A$, then compression results in an internal torque density. However, rotation of the material induces no stresses. The black arrows represent applied strains, the dotted lines denote the material's undeformed shape, and the blue icons symbolize the internal stresses. 
 {\bf b.~}The odd elastic modulus $K^o$ couples the two independent shear deformations. In contrast to passive anisotropic solids, the induced stress is always rotated 45$^\circ$ counter-clockwise with respect to the applied strain. 
 {\bf c.~}A small patch of odd elastic material is subjected to a closed cycle in strain space. Initially, the solid undergoes a counter-clockwise rotation through angle $\epsilon_\theta$. Then the solid undergoes volumetric strain $\epsilon_V$, inducing a torque density $A \epsilon_V$. Next, as the object is rotated clockwise through angle $\epsilon_\theta$, the solid does work $A \epsilon_V \epsilon_\theta$ on its surrounding. Finally, the object is compressed to its original size. As the initial and final configurations are identical, zero net work is done due to bulk modulus $B$. The total work evaluates to $A$ times the area enclosed in deformation space: $ \epsilon_V \epsilon_\theta$. {\bf d.~}An analogous cycle involving only shear stress and shear strain. The horizontal axis ($S_1$) denotes shear with extension along the horizontal, and the vertical axis ($S_2$) represents shear at 45$^\circ$. The work done is $2K^o$ times the area in deformation space. {\bf e-g.~}A network of metabeams that leads to odd elasticity. The essential feature of the metabeam is a chiral torque proportional to compression or extension. A metabeam can be subjected to a cycle of displacements that extracts work. The work done is proportional to area enclosed times odd spring constant $k^o$. For this example, the spring does net work during the third leg of the cycle (pink to purple).
}
 \label{fig1}
 \end{figure*}

The presence of $A$ and $K^o$ violates a basic symmetry of the elastic tensor: $K_{ijmn}=K_{mnij}$. 
This so-called \emph{major symmetry} stems from the assumption that the stresses are gradients of a free energy $f=\frac{1}{2}K_{ijmn}u_{ij}u_{mn}$, see Methods. In this Letter, we lift this assumption and consider an additional contribution to the stiffness tensor absent in previous hydrodynamic theories of active matter~\cite{Marchetti2013,Lau2003, Furthauer2013a,Hemingway2015,Murrell2015,Prost2015,Bi2016,vanZuiden2016,Needleman2017,Henkes2019,Woodhouse2018,Julicher2018,Maitra2018}: $K_{ijmn}=K_{ijmn}^e+K_{ijmn}^o$ with $K_{ijmn}^o=-K_{mnij}^o$, which is antisymmetric or odd under exchange of a pair of indices, in addition to the symmetric or even component $K_{ijmn}^e=K_{mnij}^e$. 
When specialized to two-dimensional isotropic media, $K_{ijmn}^o$ gives rise to two additional elastic moduli $K^o$ and $A$, forbidden by energy conservation, but allowed in active media and metamaterials with non-conservative interactions.
We stress that odd elasticity is a property of solids and is a distinct phenomenon from odd or Hall {\it viscosity} and related effects~\cite{Avron1995,Avron1998,Wiegmann2014,Banerjee2017,Soni2018} that pertain to the transport properties of fluids.

Since odd elasticity cannot be obtained from a free energy, the presence of $A$ and $K^o$ entails the ability to take an odd elastic medium through a closed cycle of quasistatic deformations with non-zero total work $\Delta w = \oint \! K_{ijmn}^o u_{mn} \dd u_{ij}$ done by (or on) the material. In Fig.~1c, we show such a cycle in the space of rotations and dilations. The initial and final configurations are identical, hence, zero work is done by the conservative part $K_{ijmn}^e$. 
By contrast, the total work done due to the odd contribution $K^o_{ijmn}$ is equal to the modulus $A$ times the area enclosed by the cycle in the space of deformations. Fig.~1d shows an analogous cycle which involves only shear stress and shear strain. This cycle does not require any torque density, hence it can be operated in a solid for which $A=0$. In this case, work is proportional to the elastic coefficient $K^o$ times the area in deformation space. Note that in both cycles the crucial feature is nonreciprocity~\cite{Fleury2014,Coulais2017}: compression induces torque, but rotation does not induce pressure; similarly, $S_2$ strain induces positive $S_1$ stress, but $S_1$ strain induces negative $S_2$ stress. If the cycle is performed in reverse, $\Delta w$ switches sign. The elastic energy cycle is local: different portions of an extended medium can operate independent cycles. Moreover, each material patch can operate two cycles simultaneously, one in the space of compression/rotation, and one in the space of shear strains.

Since energy can be extracted from an infinitesimal patch of the odd elastic solid, the microscopic constituents comprising the material must be active. For simplicity, we assume that the solid is made of particles that interact via a non-conservative pairwise force law $ \vb F ( \vb u)$, which depends only on the relative displacement $ \vb u$ from equilibrium. 
In the linear approximation, the most general expression for such a force law reads: 
\begin{equation}
\vb F (\vb u) = (-  k {\hat{\vb  r}} + k^o {\skew{2}\hat{\boldsymbol {\phi}}})   \,\, \vb u \cdot \hat{\vb r},
\label{eq2}
\end{equation}
where  ${\hat{\vb  r}}$ denotes the unit vector along the bond orientation and ${\skew{2}\hat{\boldsymbol {\phi}}}$ is the unit vector perpendicular to the bond. Consequently, for two-body interactions, the minimal necessary ingredient is that extension (compression) of a bond results in a clockwise (counterclockwise) torque generated by the transverse forces, $F$, represented by red arrows in Fig.~1f. 
The parameter $k$ sets the strength of the spring potential, and $k^o$ sets the strength of the transverse, nonconservative force~\footnote{The tangential component of force need not vanish when the bond is at its rest length (e.g., in a crystal of fluid vortices). So long as the tangential force permits a nonzero linear term in the Taylor expansion about an equilibrium length, one can obtain the same transverse response.}.

In the S.I., we show that a triangular lattice of particles interacting with these non-conservative bonds has odd moduli given by $A = 2K^o= \frac{\sqrt 3}{2} k^o$.   
Moreover, odd elastic solids can exist with zero net torque density. In the S.I., we show that the bond strengths in a honeycomb lattice with nearest-neighbor (NN) and next-nearest-neighbor (NNN) interactions can be tuned so as to eliminate the net torque within each of the hexagonal plaquettes (shown in S.I.~Fig.~S1). The result is a material in which $A = 0$, but $K^o= \frac{\sqrt 3}{2} k_2^o$, where $k_2^o$ refers to the NNN-bond strength.
We stress that any bonds exerting forces for which $\nabla \times \vb F (\vb u) \neq 0$ 
will generically lead to odd elastic moduli in the linear approximation~\footnote{Note that friction cannot be written as a function of position alone.}. Moreover, if a single constituent such as an active bond can perform work during a quasi-static cyclic deformation (see S.I. and Fig.~1g), odd elasticity will generically arise irrespective of the microscopic details. Complex mechano-chemical interactions provide additional examples of non-conservative bonds.

\begin{figure}[t!]
 \includegraphics{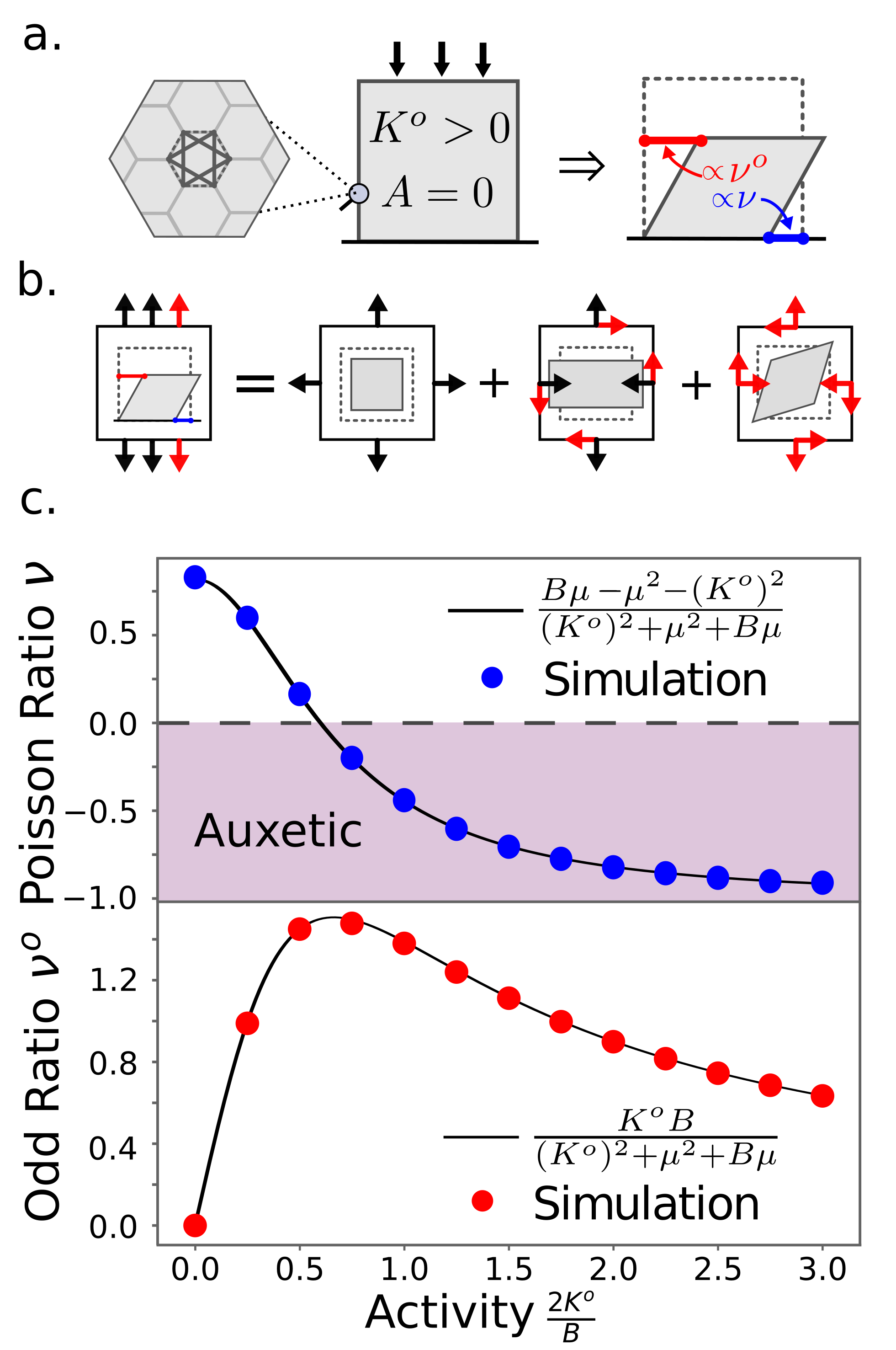}
 \caption{{\bf Statics in an odd elastic solid.} 
 {\bf a.~} 
 A honeycomb lattice with nearest-neighbor and next-nearest-neighbor odd springs can have $K^o>0$ and $A=0$ (and $B, \mu >0$). When subject to uniaxial compression, such a solid responds by both net contraction [proportional to $\nu$ (blue)] and horizontal deflection [proportional to $\nu^o$ (red)].
 {\bf b.~}Force balance in the uniaxial compression is shown schematically. Net strain can be decomposed into compression
 and shear in two directions. The resulting boundary stresses (arrows) cancel pressure on top and bottom surfaces and maintain no stress on the sides. Black arrows show the response in the absence of odd elasticity, while the red arrows show stress $S_2$ due to nonzero $K^o$, which causes simple shear strain ($S_1$). In turn, $S_1$ generates a negative pressure on the free sides, which causes the system to horizontally contract.  {\bf c.~}Analytical calculations for odd and Poisson ratios with numerical validation. Simulations are performed using the honeycomb lattice, see S.I.}
 \label{fig2}
 \end{figure}

\begin{figure*}[t!]
\includegraphics{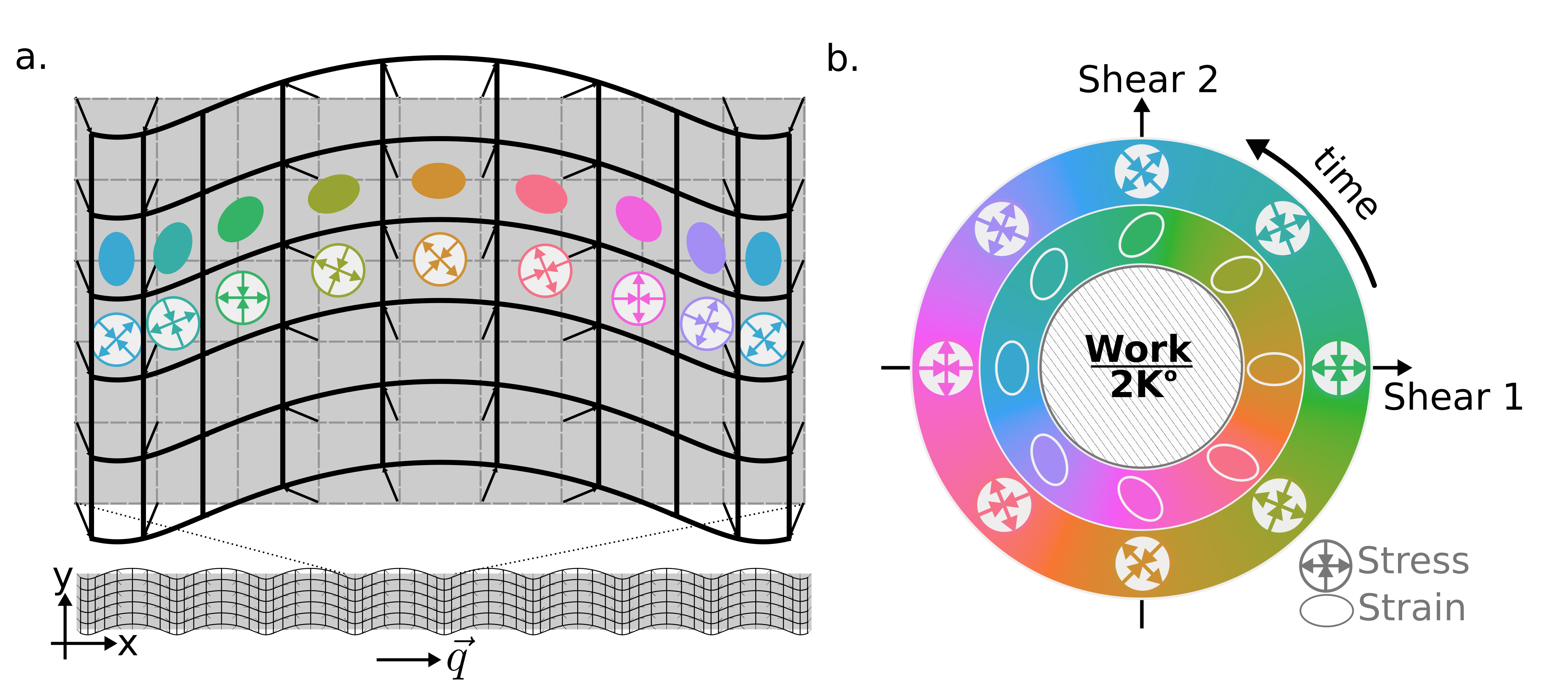}
\caption{{\bf Active elastic waves.} {\bf a.~}Real-space profile of an overdamped odd elastic wave traveling in the positive $x$-direction (for $K^o \gg A, B, \mu$). The light-grey background shows the undeformed material: the wave deforms the background grid into the thick black mesh. The ellipses illustrate the shear strain in a material patch and the disk-confined arrows represent the local shear stress. {\bf b.~}If a single material patch is tracked in time, the strain in the material traces out a circle in shear space. This circular trajectory encloses an area in strain space such that internal energy balances dissipative losses. The other essential ingredient for wave propagation is that stress and strain inside each patch are 90$^\circ$ out of phase (color represents time). (See Supplementary Movie 1.)}
\label{fig3}
\end{figure*}

\begin{figure}[t!]
\includegraphics{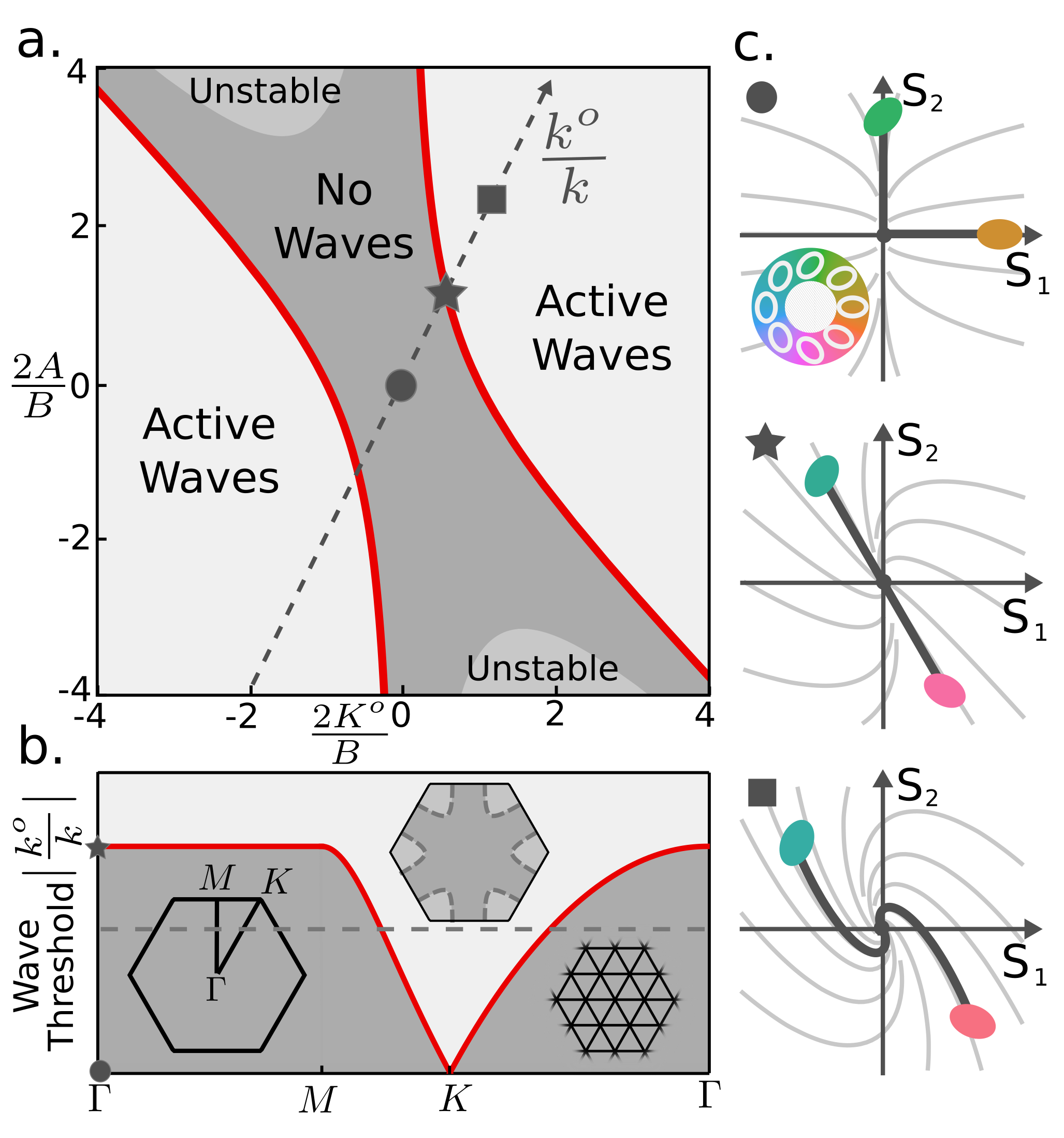}
\caption{{\bf Phase diagram and exceptional points for odd elastic waves.} {\bf a.~}Phase diagram for waves in an overdamped odd elastic solid. Red curves represent the boundary outside of which active waves can be sustained. {\bf b.~}A cut ($\Gamma M K \Gamma$) through the space of wavevectors (first Brillioun zone) of a triangular lattice with generalized Hookean springs. The microscopic activity in the springs is characterized by the ratio $\abs{\frac{k^o}{k}}$ between odd spring constant $k^o$ and conservative spring constant $k$. The threshold for active waves varies across the Brilloin zone, with the elastic limit describing the region near $\Gamma$. 
The middle inset shows the regions of the Brillouin zone (light grey) in which waves propagate (for $\abs{\frac{k^o}{k}}$ corresponding to the horizontal dashed line). 
{\bf c.~}The eigenmodes for three relative values of the elastic moduli, showing trajectories in shear space ($S_1$ and $S_2$, c.f. Fig.~3). At zero activity 
(\protect\markercirc), the modes correspond to longitudinal and transverse waves, whose eigenvectors are orthogonal in $S_1$-$S_2$ space. At the exceptional point 
(\protect\markerstar), the eigenmodes become colinear. Above the exceptional point 
(\protect\markersq), the eigenmodes acquire a circular polarization, performing a spiral through simultaneous rotation and attenuation in phase space. (See Supplementary Movie 3.)}
\label{fig4}
\end{figure}

When the odd moduli $A$ and $K^o$ are present, even the most familiar elastic phenomena appear in a new guise. To elucidate the role of odd shear coupling $K^o$, we depict in Fig.~2a the uniaxial compression of an odd elastic material having $K^o, B, \mu > 0$ and $A =0$. In passive elastostatics, uniaxial compression is used to determine the Poisson ratio $\nu \equiv -\frac{u_{xx}}{u_{yy}}$, i.e., the ratio between horizontal strain $u_{xx}$ and vertical strain $u_{yy}$. For passive solids, the Poisson ratio can be made negative by 
tuning lattice geometry or using three-body interactions~\cite{Bertoldi2010}. Here, we focus solely on the effect of activity. We find that, regardless of microscopic realization, increasing the activity $\abs{\frac{2K^o}{B}}$ pushes the solid towards the auxetic limit of $\nu = -1$. Moreover, an additional response, not observed in passive elasticity, emerges: the odd solid exhibits a horizontal deflection of the top surface with respect to the bottom surface, which we quantify via the odd ratio: $\nu^0 \equiv - \frac{u_{yx}}{2u_{yy}}$.  Whereas in passive isotropic solids, the odd ratio is zero due to left-right symmetry, the odd shear coupling $K^o$ manifestly breaks chiral symmetry and thus allows the deflection. In Fig.~2b, we illustrate the auxetic behavior and show how the odd ratio results from force balance at the boundary. 
In Fig.~2d-e, we plot analytical predictions for $\nu$ and $\nu^o$ as solid black lines. To validate our analytical results, we simulate a honeycomb lattice that has both NN and NNN bonds. Using an analytic coarse-graining procedure (see S.I.), we obtain the desired values of $K^o$, $\mu$, $B$, and $A$ from the microscopic spring constants. The measured Poisson ratio, plotted in Fig.~2d, agrees well with the prediction of the continuum theory without any fitting parameters.

We now turn to odd elastodynamics and study wave propagation in the overdamped regime in which energy injection due to activity can counteract dissipation~\cite{Toner2005,Marchetti2013,Geyer2018,Soni2018}. A distinctive feature of odd elastic waves is that they exist even when the bulk and shear moduli are vanishingly small. Fig.~3a shows a snapshot of a plane wave traveling to the right in an overdamped solid in which $K^o \gg A, B, \mu$. (See Supplementary Movies 1 and 3.)
The colored ellipses represent the strain in regions bounded by the thick, black lines (c.f., the ellipses in Fig.~1c). In the row underneath the ellipses, we show the shear stress. 
In Fig.~3b, we plot the stress and strain of a single deformed square as a function of time (indicated by color) in the space of shear $S_1$ and $S_2$.
Shear stress and shear strain encode the basic mechanism for overdamped elastic wave propagation. 

In the overdamped regime without odd elasticity, stress, strain, and momentum are all in phase and no elastic wave propagation can exist. 
For an overdamped odd elastic solid, Fig.~3b illustrates two crucial features. First, stress and strain are out of phase due to the antisymmetric shear coupling $K^o$. Since stress and velocity are in phase for an overdamped wave, an odd elastic solid can mimic the phase delay between strain and velocity that enables wave propagation in inertial, passive solids. 
Second, the trajectory of the wave in strain space traces out a circle. 
This circle indicates the emergence of an autonomous, self-sustaining elastic engine cycle, in which the system converts internal energy into mechanical work to offset dissipative losses [c.f. Fig.~1c]. For a wave of amplitude $R$ and wavenumber $q$, the circle has radius $qR$. Therefore,
by balancing the dissipative losses $2 \pi \eta \omega R^2$ in each period against energy injected $2\pi K^o q^2 R^2$, we arrive at the speed of sound for these dispersive waves, given by the group velocity $d\omega/dq = 2K^o q/\eta$ (see S.I.).
Figure~3 shows active-wave propagation in the regime dominated by $K^o$. 
When $B$ and $\mu$ are nonzero, the waves decay exponentially with a rate proportional to $\mu + B/2$.

As activity decreases, there is a sharp cutoff below which active waves can no longer be sustained. 
Figure~4a shows this threshold for active wave propagation, highlighted in red. The phase diagram in Fig.~4a summarizes the dynamic behavior of isotropic odd elastic solids, regardless of their microscopic realization. To understand the spectrum at shorter wavelengths, a microscopic structure must be specified. In Fig.~4b, we consider an unbounded triangular lattice of springs with conservative spring constant $k$ and odd spring constant $k^o$. 
Analytic coarse graining shows that this microscopic realization corresponds to a position (set by $k^o/k$) on the dashed line in Fig.~4a. For large activity (i.e., $\abs{k^o/k}$), elastic waves propagate, but at the critical value $\abs{k^o/k} = \frac1{\sqrt 3}$, these waves disappear. Elasticity describes the dynamics in the neighborhood of $\Gamma$, and the $\Gamma M K \Gamma$ cut in Fig.~4b shows how the wave-propagation threshold varies depending on the wavevector within the Brillioun zone. 
The middle inset of Fig.~4b highlights the regions in the Brillioun zone (light grey) for which waves can propagate when, as an example, $\abs{\frac{k^o}{k}}$ is given by the horizontal dashed line.
The surprising feature is the existence of waves at short lengthscales, well below the critical value in the continuum theory of Fig.~4a. Figure~4c and Supplementary Movie 3 illustrate the onset of elastic waves using three cases. In the absence of activity (\markercirc), the two eigenmodes are longitudinal and transverse. As activity increases, 
the eigenvectors are no longer orthogonal, and at the threshold $k^o/k = \frac1{\sqrt3}$, the eigenvectors are co-linear (\markerstar). The singularity caused by the degeneracy of the eigenvectors is known as an exceptional point~\cite{Bender1998,Heiss2012}. Above the exceptional point (\markersq), odd elastic waves propagate with circular polarization, tracing out a spiral in shear space due to attenuation.
In the limit $\frac{k^o}{k}\gg 1$, the waves become self-sustaining and the spiral expands into an ellipse.

In summary, our work brings to light a hitherto neglected facet of elasticity that applies generically to systems for which elastic energy cannot be defined. Future work will explore applications of our theoretical framework to biomechanical systems~\cite{Foster2015,Needleman2017}, kinematics of systems with transverse interactions such as gyroscopes or vortex lattices~\cite{Nash2015}, exotic viscoelastic quantum Hall states~\cite{Offertaler2019} and active metamaterials functioning as emergent soft robots that harvest energy, transmit it using odd mechanical waves, and perform work at designated sites.

\textbf{Acknowledgments} 
A.S., W.T.M.I., and V.V.~acknowledge primary support through the Chicago MRSEC, funded by the NSF through grant DMR-1420709. C.S.~was supported by the National Science Foundation Graduate Research Fellowship under Grant No. 1746045. W.T.M.I.~acknowledges support from NSF EFRI NewLAW grant 1741685. P.S.~was supported by the Deutsche Forschungsgemeinschaft via the Leibniz Program.

\section{Methods}
\subsection{Elastic energy and
symmetries of the stiffness tensor}
The standard theory of elasticity begins with the postulation of an elastic free energy density $f$ (see e.g., Ref.~\cite{Landau7}). 
The requirement that the free energy be invariant under translations of the solid implies $\pdv{f}{u_i}=0$, so the free energy is only a function of gradients of $u_j$. In the limit of long-wavelength deformations, the lowest-order gradient $u_{ij} = \partial_i u_j$ dominates. Mechanical stability implies $\eval{\pdv{f}{u_{ij}}}_{u_{ij =0}}=0$, so the lowest order term in strain must be quadratic. 
To linear order, the distances between points change only due to changes in the symmetrized displacement gradients $u^s_{ij} \equiv \frac{1}{2} (\partial_i u_j + \partial_j u_i)$ (see, e.g., Ref.~\cite{WarnerTerentjev}). Therefore, $u^s_{ij}$ defines the linear strain tensor.
Thus, the elastic free energy may be written as:
\begin{equation}
   f=\frac{1}{2}C_{ijmn}u^s_{ij}u^s_{mn},
   \label{eq:f}
\end{equation}
where $C_{ijmn}$ is a constant rank-4 tensor. 

The stress tensor is given by:
\begin{equation}
    \sigma^{\mathrm{eq}}_{ij} = \pdv{f}{u^s_{ij}} 
    = \frac12 \qty(C_{ijmn} + C_{mnij} ) u^s_{mn}. \label{Ceq}
\end{equation}
Thus, we obtain the constitutive relation $\sigma^{\mathrm{eq}}_{ij} = K_{ijmn} u_{mn} $, where $K_{ijmn}$ is known as the elastic, or stiffness, tensor. From Eq.~(\ref{Ceq}) we see that 
\begin{equation}
K_{ijmn}=\frac12 \qty(C_{ijmn} + C_{mnij} ) = K_{mnij}.
\end{equation}
Therefore, if a solid medium obeys a linear constitutive relation which follows from a free energy, then the elastic tensor must obey the major symmetry $K_{ijmn} = K_{mnij}$. 
Note that the definition $\sigma^{\mathrm{eq}}_{ij} \equiv \partial f/ \partial u^s_{ij}$ implies that the stress is symmetric, $\sigma^{\mathrm{eq}}_{ij} = \sigma^{\mathrm{eq}}_{ji}$ (because $u^s_{ij}$ is symmetric). In turn, this means that the non-active solid has no internal torques (evaluated as $\sigma^{\mathrm{eq}}_{ij} \epsilon_{ij} = 0$, where $\epsilon_{ij}$ is the two-dimensional Levi-Civita symbol).

In order to consider an odd elastic component $K^o_{ijmn} = - K^o_{mnij}$,
we cannot start in the usual way from an elastic free energy. Instead, we begin from the constitutive relations directly: $\sigma_{ij} = K_{ijmn} u_{mn}$.
If, unlike Eq.~(\ref{Ceq}), the constitutive relations are not derived from an elastic free energy density, then an odd elastic component
can exist.

\subsection{Classification of elastic moduli} 
\noindent We suppose a solid body undergoes a deformation such that a point originally located at position $\vb x$ (having components $x_i$) ends up at location $X_i(\vb x)$. We define the displacement vector field for the solid to be $u_i(\vb x) \equiv X_i(\vb x) - x_i$, and define the displacement gradient tensor to be $u_{ij} (\vb x) \equiv \partial_i u_j (\vb x) $ (i.e., $u_{ij}$ is related to the deformation gradient tensor $\Lambda_{ij} \equiv \partial X_i(\vb x)/\partial x_j$ via $u_{ij} = \Lambda_{ij} - \delta_{ij}$, where $\delta_{ij}$ is the Kronecker-$\delta$). 
Note that to linear order, $u_{ij}$ plays the role of an unsymmetrized elastic strain tensor, which under the assumptions of rotational invariance can be symmetrized in the usual way (see below).
The continuum version of Hooke's law postulates that if the strains (i.e., displacement gradients) are sufficiently small, the stress field $\sigma_{ij} (\vb x)$ induced in a solid due to the strain is given by:
\begin{align}
    \sigma_{ij} (\vb x) = K_{ij mn} u_{mn} (\vb x), \label{elastdef}
\end{align}
where $K_{ijmn}$ is known as the elastic tensor or the stiffness tensor. This assumption underlies  linear elasticity theory. In what follows, we assume that the material is homogeneous, i.e., that $K_{ijmn}$ is constant in space. The components of $K_{ijmn}$ are known as elastic moduli, and they are the coefficients of proportionality between stress and strain that characterize the elastic behavior of a solid.  

As we now show, basic assumptions about forces within the solid, such as conservation of angular momentum and conservation of energy, guarantee symmetries for the form of the elastic tensor. For convenience, we work in two dimensions and we introduce the following basis for $2\times2$ matrices:
\begin{align}
    \tau^0 &= \mqty(1 & 0 \\ 0 & 1) \\
    \tau^1 &= \mqty(0 & -1 \\ 1 & 0) \\
    \tau^2 &= \mqty(1 & 0 \\ 0 & -1) \\
    \tau^3 &= \mqty(0 & 1 \\ 1 & 0).
\end{align}
In this basis, we define:
\begin{align}
    u^0(\vb x) &= \tau^0_{ij} u_{ij} (\vb x)  && \qq*{Dilation}  \\
    u^1(\vb x) &= \tau^1_{ij} u_{ij} (\vb x) && \qq*{Rotation}\\
    u^2(\vb x) &= \tau^2_{ij} u_{ij} (\vb x) && \qq*{Shear strain 1} \\
    u^3(\vb x) &= \tau^3_{ij} u_{ij} (\vb x) && \qq*{Shear strain 2}
\end{align}
These four independent components define the full displacement gradient tensor and can be interpreted as follows: $u^0$ measures the local, isotropic dilation of the solid. A dilation corresponds to change in area without change in shape or orientation; $u^1$ measures the local rotation, which corresponds to change in orientation without change in shape or area (Under transformations of 2D space, $u^0$ has the symmetry of a scalar and $u^1$ has the symmetry of a pseudo-scalar.) The two components $u^2$ and $u^3$ define the shear strain, which corresponds to change in shape without change in area or orientation. (Under rotations of 2D space, $u^2$ and $u^3$ both behave as bivectors, i.e., double-headed arrows. The combined space spanned by $\tau^2$ and $\tau^3$ is precisely that of symmetric traceless tensors.) Specifically, $u^2$ measures shear strain with extension along the $x$-axis and contraction along the $y$-axis (or vice versa), which we dub shear-1 for convenience. One the other hand, $u^3$ measures shear-2, which has the axis of extension rotated $45^\circ$ counter-clockwise with respect to shear-1. Note that two independent shear vectors (in addition to compression and rotation) are needed to form a complete basis for arbitrary deformations. 

We choose the same basis for the stress tensor:
\begin{align}
    \sigma^0(\vb x) &= \tau^0_{ij} \sigma_{ij} (\vb x)  && \qq*{Pressure}  \\
    \sigma^1(\vb x) &= \tau^1_{ij} \sigma_{ij} (\vb x) && \qq*{Torque density}\\
    \sigma^2(\vb x) &= \tau^2_{ij} \sigma_{ij} (\vb x) && \qq*{Shear stress 1}\\
    \sigma^3(\vb x) &= \tau^3_{ij} \sigma_{ij} (\vb x) && \qq*{Shear stress 2}
\end{align}
The physical interpretation of these stresses are analogous to the strains: $\sigma^0$ is the (negative) of the isotropic pressure. The component $\sigma^1$ captures the antisymmetric part of the stress, i.e., the torque density. The two remaining components, $\sigma^2$ and $\sigma^3$, correspond to shear stresses. 

In this notation, we express the elastic tensor as a $4\times4$ matrix $K^{\alpha \beta} = (\tau^\beta)^{-1}_{ij} K_{ijmn} \tau^\alpha_{mn}$. Then Eq.~(\ref{elastdef}) becomes: 
\begin{align}
\mqty( \sigma^0 (\vb x) \\ \sigma^1(\vb x) \\ \sigma^2(\vb x) \\ \sigma^3(\vb x) ) = & 2
\mqty(
K^{00} & K^{01} & K^{02} & K^{03 }\\ 
K^{10} & K^{11} & K^{12} & K^{13 } \\
K^{20} & K^{21} & K^{22} & K^{23 } \\
K^{30} & K^{31} & K^{32} & K^{33 } \\
)
\mqty( u^0 (\vb x) \\ u^1(\vb x) \\ u^2(\vb x) \\ u^3(\vb x) ). \label{Kab}
\end{align}

Here, we review how the assumptions of symmetry and conservation laws in the standard theory of elasticity constrain the form of $K^{\alpha \beta}$:
\begin{itemize}
\item[] {\bf Assumption 1: Deformation dependence (DD).} A solid-body rotation of a material does not change the distance between points within that material (i.e., the metric). Therefore, one generally assumes that solid-body rotations do not induce stress, because stresses should only emerge if the object is deformed, not merely due to changes in orientation.  The assumptions that solid-body rotations do not induce stress is equivalent to the minor symmetry $K_{ijmn} = K_{ijnm}$, or in the notation of Eq.~(\ref{Kab}), $K^{\alpha 1} =0$ for all $\alpha$.  Note that in our derivation, we use the displacement gradient tensor $u_{ij} \equiv \partial_i u_j$ instead of the linear symmetrized strain $u^{s}_{ij} \equiv \frac12 \qty(\partial_i u_j + \partial_j u_i)$ or the full nonlinear strain tensor $u^{nl}_{ij} \equiv \frac12 (\Lambda_{ik}\Lambda_{kj} - \delta_{ij})$. The full tensor  $u^{nl}_{ij}$ is rotationally invariant at all orders, and at linear order reduces to $u^{s}_{ij}$ (see, e.g., Ref.~\cite{WarnerTerentjev}). If $K_{ijmn}$ has the minor symmetry $K_{ijmn} = K_{ijnm}$, then the product $K_{ijmn}u_{mn}$ is the same whether or not $u_{mn}$ is symmetrized. We choose to work with the displacement gradient tensor $u_{mn}$ (i.e., unsymmetrized strain) to be explicit about the assumption of non-coupling to rotation.

    \item[] {\bf Assumption 2: Isotropy (IS).} Isotropy implies that the elastic tensor remains unchanged through a rotation of the coordinate system. A passive rotation of the coordinate system through an angle $\theta$ maps $K^{\alpha \beta} \mapsto R^{\alpha \gamma}(\theta) K^{\gamma \sigma} R^{\beta \sigma} (\theta) $, where
    \begin{align}
        R^{\gamma \sigma} (\theta)= 
        \mqty( 
        1 & 0 & 0 & 0  \\
        0 & 1 & 0 & 0 \\
        0 & 0 & \cos(2 \theta) & \sin(2 \theta) \\
        0 & 0 & -\sin(2 \theta) & \cos(2 \theta) \\
        ).
    \end{align}
    The requirement of isotropy can be restated as $K^{\alpha \beta} = R^{\alpha \gamma}(\theta) K^{\gamma \sigma} R^{\beta \sigma} (\theta) $ for all $\theta$. Hence, under the assumption of isotropy, the most general form of the elastic tensor is:
    \begin{align}
       K^{\alpha \beta}= 2\mqty(
K^{00} & K^{01} & 0 & 0\\ 
K^{10} & K^{11} & 0 & 0 \\
0 & 0 & K^{22} & K^{23} \\
0 & 0 & -K^{23} & K^{22} \\
).
    \end{align}

    \item[] {\bf Assumption 3:  Conservation of energy (CE).} In Section A of the S.I., we show that an an elastic tensor is compatible with the conservation of energy if and only if $K_{ijmn}= K_{mnij}$. In the notation of Eq.~(\ref{Kab}), the condition for energy conservation is $K^{\alpha \beta}=K^{\beta \alpha} $. 
    
    \item[] { \bf Assumption 4: Conservation of angular momentum (CAM).} A material conserves angular momentum if it has no internal sources of torque. In this case, one requires that $\sigma_{ij} = \sigma_{ji}$, or equivalently $\sigma^1(\vb x) =0$. To impose this constraint, one has to impose the first minor symmetry for the elastic tensor $K_{ijmn} = K_{ji mn}$, or in the notation of Eq.~(\ref{Kab}), angular momentum conservation corresponds to $K^{1\alpha} =0$ for all $\alpha$. Notice that a material need not conserve angular momentum. For example, a material composed of spinning parts which accelerate or decelerate based on an internal actuation mechanism can have an internal source of angular momentum via local torques. 
\end{itemize}

Note that if assumption 1 (deformation dependence) is the only assumption present, then $K^{\alpha \beta}$ has $12$ independent components. In the standard theory of linear elasticity with energy conservation (assumptions 3), the number of independent components is reduced to $6$.  Note that assumptions 1 and 3 together imply the minor symmetry imposed by assumption 2 with no additional restrictions. If one further assumes isotropy, the form of the elastic tensor is restricted to have 2 independent components $B$ and $\mu$:
\begin{align}
       K^{\alpha \beta}= 2\mqty(
B & 0 & 0 & 0\\ 
0 & 0 & 0 & 0 \\
0 & 0 & \mu & 0 \\
0 & 0 & 0 & \mu \\
). \label{isoform}
    \end{align}
Here, $B$ is the familiar bulk modulus, which is the proportionality constant between compression and pressure; $\mu$ is the shear modulus, which is the proportionality constant between shear stress and shear strain.

In this work, we retain only assumptions 1 and 2 (deformation dependence and isotropy). We assume deformation dependence because in the solids we consider stress only arises as a result of relative displacements (i.e., changes in the material's metric), which immediately implies assumption 1.
Note that isotropy is not a strict requirement, and many crystalline solids have anisotropic stiffness tensors. However, we consider only the isotropic case for simplicity.
In this work we study odd elasticity, which arises when we lift assumption 3 (conservation of energy). Assuming isotropy, the general form of the elastic tensor under these relaxed assumptions is
\begin{align}
       K^{\alpha \beta}= 2\mqty(
B & 0 & 0 & 0\\ 
A & 0 & 0 & 0 \\
0 & 0 & \mu & K^0 \\
0 & 0 & -K^0 & \mu \\
). \label{isoform2}
    \end{align}
In this case, there are two new moduli: $A$ and $K^o$. As described in the text, $A$ couples compression to internal torque density. The modulus $K^o$, like the shear modulus $\mu$, is a proportionality constant between shear stress and shear strain. However, $K^o$ mixes the two independent shears in an anti-symmetric way. 

In our work, assumption 3 (energy conservation) is independent of assumption 4 (angular-momentum conservation). This implies that $K^o$ and $A$ are independent elastic moduli. We consider both cases: case (i), in which angular momentum is conserved and the solid has no internal torque density (i.e., assumption 4 holds and $A = 0$) as well as case (ii) in which internal torques and odd elasticity coexist (i.e., assumption 4 does not hold and $A \ne 0$). 
Even if $A = 0$, the modulus $K^o$ can be nonzero. Hence, the existence of odd elasticity is not contingent on the presence of antisymmetric stress (or, equivalently, local active torques). 

In index notation, the most general form of the elastic tensor from Eq.~(\ref{isoform2}) is:
\begin{align}
    K_{ijmn} &= B \delta_{ij}\delta_{mn} + \mu \qty(\delta_{in}\delta_{jm}+\delta_{im}\delta_{jn}-\delta_{ij}\delta_{mn}) \nonumber \\
    &+ K^o E_{ijmn} -A \epsilon_{ij}\delta_{mn}, \label{Kgeneral} \nonumber 
\end{align}
where 
\begin{align}
    E_{ijmn}&\equiv \frac12\qty(\epsilon_{im}\delta_{jn}+ \epsilon_{in}\delta_{jm}+ \epsilon_{jm}\delta_{in}+ \epsilon_{jn}\delta_{im} ). 
\end{align}

\subsection{\bf Comparison
of odd elasticity to other approaches in active solids and metamaterials}
In this section, we compare odd elasticity to other approaches that describe complex and active solids.

Cosserat elasticity refers to a theory in which an extra microrotation field $\phi$ is introduced. In two dimensions, this pseudoscalar field captures soft rotations of a solid's microscopic constituents. Although different couplings between microrotation and elastic strain are possible, in the most common version of Cosserat elasticity, this
coupling takes the form of the term $(\phi - 
\epsilon_{ij}\partial_iu_j)^2$ added to the elastic free energy density. This Cosserat elasticity has a free energy density
\begin{equation}
    f = \frac{1}{2}K_{ijmn}u^s_{ij}u^s_{mn} + \alpha \phi^2 + \beta (\phi - 
\epsilon_{ij}\partial_iu_j)^2,\end{equation}
where $u^s_{ij} \equiv \frac{1}{2} (\partial_i u_j + \partial_j u_i)$ is the linearized strain tensor and $\alpha$ and $\beta$ are material parameters. Note the significant differences between Cosserat elasticity and odd elasticity. First, because Cosserat elasticity is based on a free energy (i.e., it describes equilibrium solids), the corresponding elastic tensor cannot allow an odd elastic component, even if the microrotation field $\phi$ is integrated out. Although microrotations allow for an off-diagonal torque-compression coupling, Cosserat solids (because they are not active) require a reactive rotation-pressure coupling, leading to a purely symmetric elastic tensor. Second, the defining feature of Cosserat elasticity is the extra microrotation field. By contrast, odd elasticity introduces no extra fields and the active terms come from stress-strain response which is disallowed in equilibrium but which is possible in an active solid.

The standard approach to many elastic systems, both near and far from equilibrium, is based on entropy generation. However, in our work the smallest objects (the constituent bonds) are macroscopic in scale. Any increases in temperature within these objects do not result in the thermal fluctuations of any translational degrees of freedom---heat flows are decoupled from the mechanics.

Other examples of solids in which extra fields are needed to describe complex elasticity include orientational order in liquid-crystal elastomers, electric fields in ferroelectrics, and structural order in shape-memory alloys. In all of these cases, the solid is not active in the way that the word is used in this work: ferroelectrics, elastomers, or shape-memory alloys do not spontaneously perform work on their environment, because these systems do not include microscopic motors that consume energy. Furthermore, the constituent microscopic bonds in these materials do not include active forces or torques and our analysis does not apply. 
Instead, these solids can be activated by applying an external field, leading to exotic mechanics.
The typical description of such activated solids relies on passive (normal) elasticity and the addition of a stress field which can generate both bulk forces and corrections to passive elasticity.
In this work, we take a different approach in which we include activity in the constituent relations via the addition of odd-elastic terms.

\pagebreak

\setcounter{equation}{0}
\setcounter{figure}{0}

\section{Supplementary Information}
\renewcommand{\thefigure}{{S\arabic{figure}}}
\renewcommand{\theequation}{S\arabic{equation}}
\subsection{\bf Elastic energy cycle}
In this section, we show that an elastic solid conserves energy if and only if $K_{ijmn} = K_{mnij}$. We represent $K_{ijmn}$ as a $4 \times 4$ matrix $K_{\alpha \beta}$ (see previous section) and write $K_{\alpha \beta} = K^e_{\alpha \beta } + K^o_{\alpha \beta}$, where $K^e_{\alpha \beta}=K^e_{\beta \alpha}$ is even (i.e., conservative) and $K^o_{\alpha \beta }= -K^o_{\beta \alpha}$ is odd (i.e., non-conservative).

The work per unit volume done on a solid in a quasi-static, infinitesimal deformation is given by
\begin{align}
    \dd w &= \sigma_{ij} \dd u_{ij}\\
    &= \frac12 \sigma^\alpha \dd u^\alpha \\
    &= \frac12 K_{\alpha \beta }u^{\alpha} \dd u^{\beta}.
\end{align}
If we take a piece of material through a path of strains that returns to the initial configuration, then the total work per unit area done on the material is:
\begin{align}
   w &=  \frac12 \oint K_{\alpha \beta }u^{\alpha} \dd u^{\beta} \\
   &= \frac12 \oint K_{\alpha \beta }^e u^{\alpha} \dd u^{\beta}+\frac12 \oint K_{\alpha \beta }^o u^{\alpha} \dd u^{\beta}.
\end{align}
Integration by parts yields:
\begin{align}
    \oint K_{\alpha \beta }^e u^{\alpha} \dd u^{\beta} &= - \oint K^e_{\alpha \beta} u^\beta \dd u^\alpha && \qq*{Integration by parts} \\
    &= - \oint K^e_{\beta \alpha} u^\alpha \dd u^\beta && \qq*{Relabel indices} \\
    &= - \oint K^e_{\alpha \beta} u^\alpha \dd u^\beta && K^e_{\alpha \beta} = K^e_{\beta \alpha}.
\end{align}
Consequently, $\frac12\oint K_{\alpha \beta }^e u^{\alpha} \dd u^{\beta} =0$. This can also be seen directly because $K_{\alpha \beta }^e$ arises from a potential energy [see first Methods section]. Because the potential energy depends only on the configuration and not on the deformation path, the energy has to be the same at the beginning and end of the closed cycle. Therefore, the contribution to net work must be zero. We now evaluate $\frac12 \oint K_{\alpha \beta }^o u^{\alpha} \dd u^{\beta}$. For an isotropic solid, the anti-symmetric part $K^o_{\alpha \beta}$ takes the form:
\begin{equation}
    K^o_{\alpha \beta} = 
    \mqty( 
    0 & -A & 0 & 0\\
    A & 0 & 0 & 0\\
    0 & 0 & 0 & 2K^o \\
    0 & 0 & -2K^o & 0
    ).
    \label{Kiso}
\end{equation}
Even in the case of a more general solid, such as one that violates isotropy or deformation dependence (see Methods), we can still choose a basis such that $K^o_{\alpha \beta}$ takes the form:
\begin{equation}
    K^o_{\alpha \beta} = 
    \mqty( 
    0 & C & 0 & 0\\
    -C & 0 & 0 & 0\\
    0 & 0 & 0 & D \\
    0 & 0 & -D & 0
    ).
    \label{K-as}
\end{equation}
Let $\{ c_0, c_1, d_0, d_1 \}$ be the basis vectors in this basis. For an isotropic solid, the basis vectors are simply 
\begin{align}
    c_0 =& \mqty(u^0 \\ 0 \\ 0  \\ 0 ), \,\, c_1 =  \mqty( 0 \\ u^1 \\ 0  \\ 0 ), \\
    d_0 =& \mqty(0 \\ 0 \\ u^2  \\ 0 ), \,\,  d_1 = \mqty( 0 \\ 0 \\ 0  \\ u^3 ).
\end{align}
The total work per unit area done on the solid can be computed by projecting the path through $4D$ strain space onto paths in the 2D subspaces of $c_i$ and $d_i$:
\begin{align}
    w = \frac{C}2 \oint \epsilon_{ij}c_j \dd c_i+ \frac{D}2 \oint \epsilon_{ij}d_j \dd d_i,
\end{align}
where in this case the $i$ and $j$ indices run over $0$ and $1$. Examples of these paths are illustrated in Fig.~1. Let $A_c$ be the region enclosed by the $c_i$ path and $A_d$ be the region enclosed by the $d_i$
path. Application of Stokes' theorem then gives:
\begin{align}
    w &= C \int_{A_c}  \dd^2 c+ D\int_{A_d}  \dd^2 d, \\
    &= C \area(A_c) + D \area(A_d).
\end{align}
To conclude, if the major symmetry $K_{ijmn}= K_{mn ij}$ holds, then the odd elastic component is zero ($K^o_{\alpha \beta}=0$), so $w =0$ and no work is done on or by the material during a closed cycle. However, if $K^o_{\alpha \beta} $ is nonzero, then a path through deformation space can always be found such that $w \neq 0$ after a closed cycle. Here, we presented the proof in two dimensions, but the same approach can be used to generalize this statement to any dimension.

\begin{figure}[t!]
    \centering
    \includegraphics{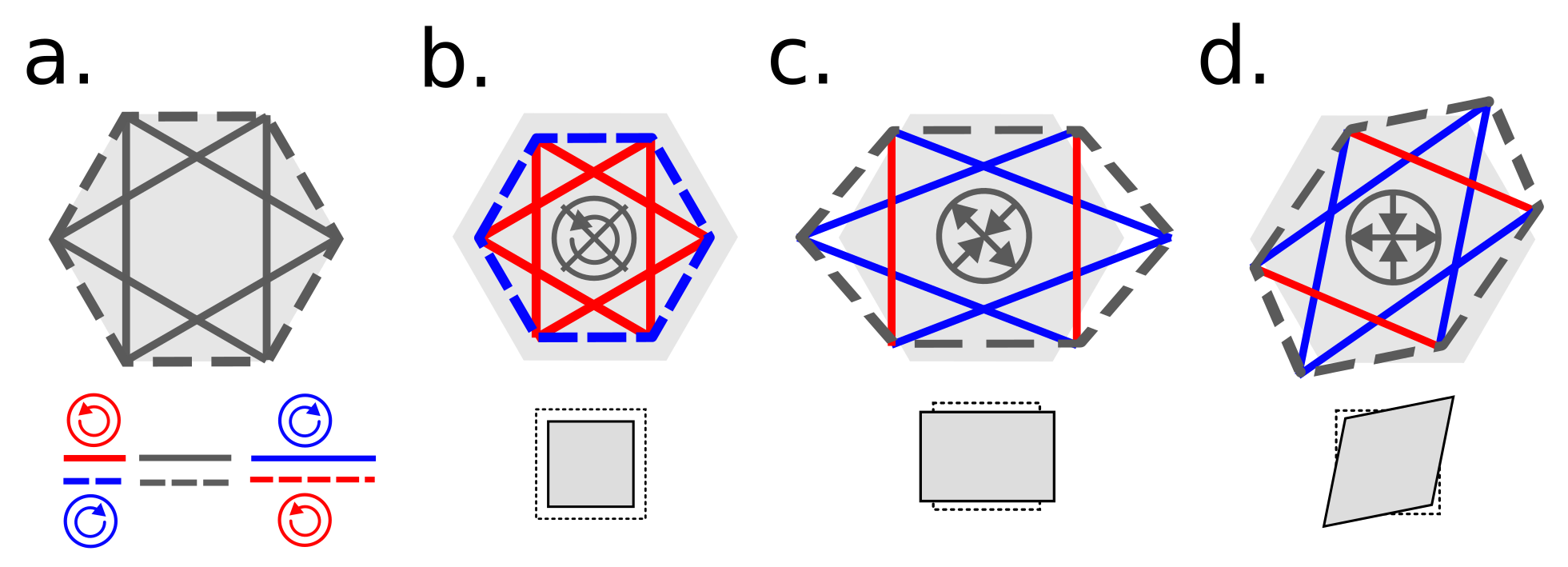}
    \caption{{\bf Torque-free honeycomb plaquette.} {\bf a.~}A honeycomb lattice can be connected out of nearest-neighbor (dashed) and next-nearest-neighbor (solid) springs with odd spring constants $k_1^o$ and $k_2^o$, respectively. The odd spring constants can be tuned so that $k^o_1 = - 6 k^o_2$, indicating that the two varieties of springs exert torques of opposite handedness when stretched or compressed. This figure illustrates qualitatively how the net internal torque density vanishes for all (linear) deformations, despite the basic units being springs which individually exert torques. {\bf b.~}A compression results in a shortening of all the springs, but the NN and NNN odd spring constants are tuned such that the net torque density is zero (blue indicates torque into the page, and red indicates torque out of the page). {\bf c.-d.~}The NN lattice is soft to shear (i.e., the bonds are unstretched), so the only contribution to the forces comes from the NNN springs. The result is a shear stress along the boundary of the unit cell with zero net torque density (the red and blue cancel). Note that the NNN bonds of a honeycomb lattice form two disjoint triangular lattices which provide shear coupling.} 
    \label{fig:Sup2}
\end{figure}

\subsection{Microscopic model}
In order to construct microscopic lattice models that exhibit odd elasticity, we introduce active springs with unusual force laws. For a spring of rest length $\ell$ that connects two particles at locations $\vb x$ and $\vb y$, the force on particle $\vb y$ as a function of the separation $\Delta \vb x = \vb y - \vb x$ is given by:
\begin{equation}
    \vb F (\Delta \vb x) = - \qty(k \frac{\Delta \vb x}{\abs{\Delta \vb x}} + k^o \frac{\Delta \vb x^*}{\abs{\Delta \vb x}} ) \qty(\abs{\Delta \vb x} - \ell), \label{Force} 
\end{equation}
where $\Delta x^*_i = \epsilon_{ij} \Delta x_j$. The first term, proportional to $k$, is the familiar Hooke's law, and the second term, proportional to $k^o$, supplies a force perpendicular to the direction of the bond or, equivalently, a torque acting on the middle of the spring along an out-of-plane axis. 

Note that Eq.~(\ref{Force}) is nonconservative since ${\nabla_{\Delta \vb x} \times \vb F \neq 0}$. Therefore, there exists cycles at the level of a single bond such that work can be extracted or injected. Hence, in the case that odd elasticity arises from pairwise interactions, the continuum elastic engine cycles actuate microscopic engine cycles at the single bond level.   

As examples, we consider this force law in the context of two different lattice geometries. First, we place the springs on a triangular lattice with lattice spacing $\ell$. The resulting moduli are 
\begin{align}
B = 2 \mu &= \frac{\sqrt 3}{2} k, \\
A = 2K^o&= \frac{\sqrt 3}{2} k^o.
\end{align}

Second, we consider a honeycomb lattice with next-nearest-neighbor springs. The nearest-neighbor springs have spring constants $k_1$ and $k_1^o$, and the next-nearest-neighbor springs have spring constants $k_2$ and $k_2^o$. The resulting moduli are:
\begin{align}
    B &= \frac{k_1+6k_2}{2 \sqrt 3} \label{HexB} \\
    A &= \frac{k^o_1 + 6k^o_2}{2 \sqrt 3} \\
    \mu &=  \frac{\sqrt3 k_2}2  \\
    K^o &= \frac{\sqrt3 k^o_2}2 \label{HexK} .
\end{align}

Note that for the honeycomb lattice, we set $k_1^o = - 6 k_2^o$. In this case, the net torque density (proportional to $A$), vanishes for all linear deformations. Importantly, the odd shear coupling does not vanish for this choice of spring constants. Thus, odd elasticity can exist without anti-symmetric stress, making theory ostensibly distinct from concepts such as Cosserat. For a mechanistic illustration of how the torque density cancels, see Fig.~S1. 

To obtain these analytically coarse-grained results, we follow the standard approach, see e.g., Ref.~\cite{Ashcroft}. Consider a lattice with $n$ particles per unit cell. For simplicity, we set the lattice spacing to 1. In the harmonic approximation, the force on each particle is given by the dynamical matrix expression:
\begin{align}
    F^{\alpha}_{i} (\vb R) = -\sum_{\vb R'} D_{ij}^{\alpha \beta} (\vb R - \vb R') u^{\beta}_{j} (\vb R'), 
    \label{dynamical}
\end{align}
where there is an implicit summation over repeated indices. The upper Greek index labels the particle in the unit cell and runs over $\alpha =0 ,\dots, n-1$, and the lower Latin index labels spatial dimension $i=x,y$. The matrix $D_{ij}^{\alpha \beta} (\vb R )$ is the dynamical matrix and is determined by the inter-particle interactions and geometry of bonds. The Fourier transform of Eq.~(\ref{dynamical}) gives: 
\begin{align}
    F^{\alpha}_{i} (\vb q) = - D_{ij}^{\alpha \beta} (\vb q) u^{\beta}_{j} (\vb q).
\end{align}
For the triangular lattice we consider, we have:
\begin{align}
    D_{ij} (\vb q) &= (k \delta_{ik} + k^o \epsilon_{ik} ) A_{kj} (\vb q) , \label{trydyn}
\end{align}
where $A_{kj}$ are components of a $2\times{}2$ symmetric matrix given by
\begin{align}
    A_{xx} (\vb q) &= 3 - 2 \cos(q_x) -\cos(\frac{q_x}2)\cos(\frac{\sqrt{3}q_y}{2}) \\
    A_{yy} (\vb q) &= 2-3\cos(\frac{q_x}2)\cos(\frac{\sqrt{3}q_y}{2}) \\
    A_{xy} (\vb q) &= A_{yx} = \sqrt{3} \sin(\frac{q_x}2) \sin(\frac{\sqrt{3}q_y}{2}).
\end{align}
For the honeycomb lattice, we find 
\begin{align}
    D^{\alpha \beta}_{ij} (\vb q) &= (k_1\delta_{ik}+k^o_1 \epsilon_{ik}) B^{\alpha \beta}_{kj} (\vb q) \nonumber \\ &+(k_2\delta_{ik}+k^o_2 \epsilon_{ik}) C^{\alpha \beta}_{kj} (\vb q),
    \label{hondyn}
\end{align}
where
\begin{align}
    B_{ij}^{00}&=B_{ij}^{11} = \frac32 \delta_{ij} \\
    B_{ij}^{01}&=(B^{10})_{ij}^\dagger \\
    &= \frac14 \mqty( -1 - e^{i q_y}-4e^{-i \qty( \frac{\sqrt3 q_x}2-\frac{q_y}2) } & \sqrt{3} \qty( 1-e^{i q_y} )   \nonumber\\ 
    \sqrt{3} \qty( 1-e^{i q_y} ) & -3\qty(1-e^{i q_y})) \\
    C_{ij}^{00}&=C_{ij}^{11}= \mqty( A_{yy} & A_{yx} \\ A_{xy} & A_{xx} ) \\
    C_{ij}^{01}&=(C^{10})_{ij}^\dagger =0.
\end{align}
If $k^o \rightarrow 0$, we recover the familiar dynamical matrices for the triangular  [from Eq.~(\ref{trydyn})] and honeycomb [from Eq.~(\ref{hondyn})] lattices. The effect of the active torques is to modify the dynamical matrix expression via the transformation $k \delta_{ik} \rightarrow k \delta_{ik} + k^o \epsilon_{ik}$. The internal active torques $k^o$ that we have introduced correspond precisely to the antisymmetric component of the dynamical matrix.

To determine the elastic tensor from the dynamical matrix, we proceed again following the standard procedure of calculating the response of a ball-and-spring lattice to large-scale deformations (see, e.g., Ref.~\cite{Lubensky2015}). A peculiarity of the approach for odd elastic systems is that it must be based on forces and constitutive stress-strain relations and not on the kinematic, potential-energy formulation. First, we perform a change of basis such that $v^{\alpha}_i = U^{\alpha \beta} u^\beta_i$, with
\begin{align}
U^{\alpha \beta }= \frac1n \mqty( 1 & 1 & 1 & 1 & \cdots & 1 \\
    -1 & n-1 & -1 & -1 &\cdots & -1 \\
    -1 & -1 & n-1 & -1 & \cdots &-1 \\
    \vdots & & & \ddots & & ).
\end{align}
Notice that $v^0_i$ is the center-of-mass coordinate. (We have assumed, for simplicity, that all the particles are of equal mass). We denote the dynamical matrix in this basis by $\tilde D_{ij}^{\alpha \beta}(\vb q)$. We use the upper-case Latin indices $A,B=1, \dots, n-1$. 

Note that $v^0_j$ can be large (compared to a lattice spacing, which we set equal to $1$), but $q_i$ and $v^A_i$ are assumed small. Furthermore, $\tilde D_{ij}^{\alpha 0}(0)=\tilde D_{ij}^{0 \beta}(0) =0$, and $\eval{\pdv{\tilde D_{ij}^{00}}{q_m}}_{\vb q =0} =0$. Therefore, expanding to lowest order in the small quantities, we find
\begin{align}
    & \mqty( i q_m \frac{V}{n} \sigma_{mi} (\vb q) \\ \ddot v^A_i (\vb q) ) \nonumber \\ &=  
    - \mqty( q_m q_n 
    \frac12 \eval{\pdv{\tilde D_{ij}^{00}}{q_m}{q_n}}_{\vb q=0}  
    & q_m \eval{\pdv{\tilde D_{ij}^{0B}}{q_m}}_{\vb q=0} \\ 
    q_n \eval{\pdv{\tilde D_{ij}^{A0}}{q_n}}_{\vb q=0} 
    &\tilde D_{ij}^{AB}(0) )
    \mqty( v^0_j(\vb q) \\ v^B_j (\vb q) )\\
    &=    - \mqty( -i q_m \frac12\eval{\pdv{\tilde D_{ij}^{00}}{q_m}{q_n}}_{\vb q=0}
    &  q_m \eval{\pdv{\tilde D_{ij}^{0B}}{q_m}}_{\vb q=0} \\ 
    -i \eval{\pdv{\tilde D_{ij}^{A0}}{q_n}}_{\vb q=0} 
    & \tilde D_{ij}^{AB}(0) )
  \mqty( u_{nj} \\ v^B_j (\vb q) ), \label{Mat1}
\end{align}
where $V$ is the (dimensionless) area of the unit cell, $u_{nj}$ are the displacement gradients, and we have used the relation $\ddot v^0_i = i q_m \sigma_{m j}/\rho$. When a macroscopic deformation is applied to the material, the microscopic unit cell deforms according to the force-balance condition $\ddot v^A_i=0$. This deformation can be non-affine if the lattice has more than one particle per unit cell. Using the force-balance condition, we write $\sigma_{mi}=K_{minj} u_{nj}$ where
\begin{equation} \label{kminj-lat}
    K_{minj} = \frac{n}V \qty(A_{minj} - T^B_{mik} \qty[C_{pk}^{AB}]^{-1} S^A_{pnj})
\end{equation}
 and 
 \begin{align}
 \mqty( A_{minj} & T_{mij}^B  \\ 
    S_{inj}^A & C^{AB}_{ij}  ) 
    \equiv \mqty(  \frac12 \eval{\pdv{\tilde D_{ij}^{00}}{q_m}{q_n}}_{\vb q=0}
    &  i \eval{\pdv{\tilde D_{ij}^{0B}}{q_m}}_{\vb q=0} \\ 
    -i \eval{\pdv{\tilde D_{ij}^{A0}}{q_n}}_{\vb q=0} 
    & \tilde D_{ij}^{AB}(0)).
\end{align}   
We assume that $C_{ij}^{AB}$ is invertible, which is equivalent to assuming that the lattice has no zero modes that preserve the center of mass of the unit cell. Additionally, we have assumed that all of the particles have equal mass. By modifying the form of $U^{\alpha \beta}$, this expression can be generalized to lattices composed of particles with different masses.

\subsection{Derivation of Poisson and odd ratios}
We consider an unbounded, odd elastic material with no net internal torque: $A=0$. In a uniaxial compression with free boundaries on the two sides parallel to the direction of compression, the material stress is captured by tensor components $\sigma_{yy}=p$, $\sigma_{xx}=\sigma_{xy}=\sigma_{yx}=0$. To fix the solid's orientation, we impose the condition $u_{xy}=0$, so that the top and bottom horizontal boundaries remain horizontal. In this elastic stability problem, we seek to solve for the strain everywhere inside the material. Using the notation $\sigma^\alpha = K^{\alpha \beta} u^{\beta}$, we obtain:
\begin{align}
    \mqty( p \\ 0 \\ -p \\ 0 ) = 
    2 \mqty( B & 0 & 0 & 0 \\
    0 & 0 & 0 & 0 \\
    0 & 0 & \mu & K^o  \\
    0 & 0 & -K^o & \mu )
    \mqty( u^0  \\ -u^3 \\ u^2 \\ u^3 ), \label{mateq}
\end{align}
Noting that $u_{xx} = \frac12 (u^0 + u^2)$ and $u_{yy}= \frac12 (u^0 - u^2)$ and $u_{yx} = \frac12 (u^3-u^1)$, we invert Eq. (\ref{mateq}) to obtain: 
\begin{align}
    u_{xx}&= \frac{p}{4}\qty(\frac{(K^o)^2+\mu^2-B\mu}{B[(K^o)^2+\mu^2]}) \label{uxx}\\
    u_{yy}&= \frac{p}{4}\qty(\frac{(K^o)^2+\mu^2+B\mu}{B[(K^o)^2+\mu^2]})\\
    u_{yx}&= \frac{p}{2}\qty(\frac{-K^o}{(K^o)^2+\mu^2}). \label{uyx}
\end{align}
From Eq.~(\ref{uxx}) through Eq.~(\ref{uyx}), we obtain expressions for the odd ratio, Poisson ratio, and Young's modulus:
\begin{align}
\nu = -\frac{u_{xx}^{s}}{u_{yy}^s} = \frac{B \mu - (K^o)^2 -\mu^2}{(K^o)^2+\mu^2+B\mu} \label{Pois} \\
\nu^o = -\frac{u_{yx}^s}{u_{yy}^s} = \frac{K^oB}{(K^o)^2+\mu^2+B\mu} \label{Oddrat}\\
E = \frac{p}{u_{yy}^s}= \frac{4B[(K^o)^2+\mu^2]}{(K^o)^2+\mu^2+B \mu} \label{young}.
\end{align}
(Recall that $u^s_{ij} \equiv \frac12(u_{ij}+u_{ji})$ is the symmetrized strain). 
This result applies exactly for an infinitely large, unbounded solid. Specifically, the result applies asymptotically in the limit of a wide solid being compressed with pressure $p$ with sliding boundary conditions on the top and bottom surfaces.

\subsection{Numerics}
In this section, we describe the molecular dynamics simulations used to validate our analytical calculations. Each particle is given a position $\vb x _i$, where the subscript $i = 1, \dots, N$ labels the particle. The particles are arranged on a lattice with a fixed bond topology. For overdamped dynamics, the physical particle positions evolve according to the equation $\gamma \dv{\vb x_i}{t} = \vb F_{i} (\{\vb x_1, \dots, \vb x_N\}(t)) $, where the right-hand side is the force on particle $i$ as computed using particle positions at time $t$. For numerical integration, we non-dimensionalize this equation by setting the lattice spacing $a = 1$ and $\frac{k_o}{\gamma} =1$, where $k_o$ is the characteristic spring constant for that lattice. This non-dimensionalization is equivalent to sending $\vb x_i \mapsto \vb x_i /a $ and $t \mapsto \frac{t \gamma}{k_o} $. 

Using this non-dimensionalization, the position of each particle is updated according to second-order Runge-Kutta. In our case,
Runge-Kutta is the appropriate integrator
because the energy is not conserved. 
In the overdamped regime, 
the governing equations of motion are first order, and this method offers an alternative to variational integrators appropriate to second-order equations of motion that conserve energy (or in which energy is nearly conserved, as in Ref.~\cite{Tsang2015}).
Using second-order Runge-Kutta, the equations
for updating the particle positions read 
\begin{align}
    \vb x_i (t+ \frac12 \Delta t) =& \frac12 \Delta t \vb F_i ( \{\vb x_1, \dots, \vb x_N\}(t) ) + \vb x_i (t) \\
    \vb x_i (t + \Delta t) =& \Delta t  \vb F_i ( \{\vb x_1, \dots, \vb x_N\}(t+\frac12 \Delta t) )+ \vb x_i (t),
\end{align}
where $\vb F_i (  \{\vb x_1, \dots, \vb x_N\}(t) ) $ is the force on particle $i$ as computed by the particle positions at time $t$. The force is given by summing Eq.~(\ref{Force}) over all the neighbors of particle $i$.  

We use this approach to simulate the following:
\begin{itemize}
    \item[]{ \bf Uniaxial compression.} The data in Fig.~2 are produced by simulating a uniaxial compression. In order to achieve $A=0$, we use a honeycomb lattice with nearest- (NN) and next-nearest-neighbor (NNN) springs. The system consists of 60 unit cells in the $x$-direction and 30 unit cells in the $y$-direction. We assign the conservative spring constants values $k_{1}=9.6 $ for NN springs and $k_{2}= 0.4$ for NNN springs (this corresponds to a solid with $B= 10\mu = 3.5$). We then set the odd spring constant $k^o_{1} = - 6 k^o_2 $, consistent with the analytical calculations for $A=0$. We perform simulations with $k^o_1$ taking values from 0 to 6.0 in increments of 0.5. The time-step is $\Delta t = 0.001 $, and the simulation runs for a total time $t_f=1$.
    
    Before we begin time-integration, we apply an initial affine strain of the form $u_{yy} = \epsilon$, $u_{xx}= -\nu \epsilon$, and $u_{yx} = - \nu^0 \epsilon $, with $\epsilon = 0.01$, where $\nu$ and $\nu^o$ are determined using Eq.~(\ref{Pois}) and Eq.~(\ref{Oddrat}), respectively, and the values of $K^o$, $A$, $\mu$, and $B$ are determined using Eqs.~(\ref{HexB}-\ref{HexK}). We then allow the system to relax using the second-order Runge-Kutta integration described above. During this relaxation, we apply a constant outward force to the top and bottom row of particles in order to induce a pressure $\epsilon E$, where $E$ is determined using Eq.~(\ref{young}). Once the system has equilibrated, we measure the strain field using least-squares regression on the particle displacements. The strain field is then used to compute the odd and Poisson ratios which appear in Fig.~2. 
    \item[]{\bf Plane waves.} In Supplementary Movie 2, we show an odd elastic plane wave propagating in the $- \hat{\vb y}$ direction through a triangular lattice of generalized Hookean springs. The springs in this simulation have spring constants $k=0$ and $k^o=1$, which correspond to coarse-grained elastic moduli $\mu = B = 0$ and $A = 2 K^o = \frac{\sqrt3}{2}$. The system is a 30$\times$30 triangular lattice with periodic boundary conditions on all sides. The particles are initially displaced according to Eq.~(\ref{eigvecs}) with displacement amplitude $0.2$, and are evolved using the second-order Runge-Kutta integration described above with time-step $\Delta t= 0.005$ and total time $t_f = 100$. 
\end{itemize}

\subsection{Elastodynamics}

In the bulk of a dissipative elastic solid, the equation of motion for the components $u_j$ of the displacement vector field reads:
\begin{equation}
\rho \ddot u_j + \eta \dot u_j = \partial_i \sigma_{ij}
    = K_{ijmn} \partial_{i} \partial_m u_n.  \label{bulkeqom:ud}
\end{equation}

In the overdamped regime, $\eta \dot u_j \gg \rho \ddot u_j$, Eq.~(\ref{bulkeqom:ud}) reduces to a first-order equation:
\begin{equation}
     \eta \dot u = \partial_i \sigma_{ij} 
    = K_{ijmn} \partial_{i} \partial_m u_n.  \label{bulkeqom}
\end{equation}
Split into components, the $K^o$- and $A$-dependent parts of the right-hand side read:
\begin{align}
     \eta \dot u_x & = K^o (\partial_x^2 + \partial_y^2) u_y - A (\partial_y^2 u_y + \partial_y\partial_x u_x),\\
     \eta \dot u_y & = - K^o (\partial_x^2 + \partial_y^2) u_x + A (\partial_x^2 u_x + \partial_x\partial_y u_y).
     \label{bulkeqomB}
\end{align}
Although Eqs.~(\ref{bulkeqom}-\ref{bulkeqomB}) appear to describe purely diffusion-like dynamics, they nevertheless allow for propagating solutions, analogous to the propagation of Avron waves in Ref.~\cite{Avron1998}.
However, whereas for Avron waves, the equation of motion describes the velocity field in a fluid, Eqs.~(\ref{bulkeqom}-\ref{bulkeqomB}) describe the displacement field within a solid.

By assuming solutions of the form $u_i(\vb x) = \tilde{u}_i(\vb q) e^{i(\vb q \cdot \vb x - \omega t)}$, we find the Fourier transform of Eq.~(\ref{bulkeqom}):
\begin{align}
    -i \omega \eta  \tilde{u}_j = q_i q_m K_{ijmn} \tilde{u}_n.
\end{align}
Using expression Eq.~(\ref{isoform2}) for $K_{ijmn}$, the right-hand side can be rewritten using a $2\times2$ matrix: 
\begin{align}
    q_i q_m K_{ijmn} \tilde u_n =  q^2 \mqty(B+ \mu & K^o \\ -K^o-A & \mu ) \mqty(u_{\parallel} \\ u_\perp), \label{matrix}
\end{align}
where $u_\parallel = \hat q_i \tilde u_i$ and $u_\perp = \epsilon_{ij} \hat q_i \tilde u_j$. The normal modes of the bulk elastic spectrum are given by the eigenvalues of Eq. (\ref{matrix}):
\begin{equation}
    \omega = -i \qty[\frac{B}2 + \mu \pm \sqrt{\qty(\frac{B}2)^2 - K^{o} A -(K^{o})^2}] \frac{q^2}\eta. \label{Spec}
\end{equation}
Eq.~(\ref{Spec}) contains information about the stability of the solid and its ability to propagate waves. For example, Fig.~4a shows boundaries for the onset of waves and instability. 
Instability occurs when the spectrum acquires a positive imaginary branch (because in that case, the exponential $e^{\Im(\omega) t}$ grows in time), which corresponds to the boundary defined by:
\begin{align}
    \tilde A = - \frac{(2 \tilde \mu + \tilde \mu^2 + (\tilde K^o)^2)}{\tilde K^o},
\end{align}
where the tilde indicates that the moduli are normalized by $\frac{B}2$.
The onset of waves occurs when $\omega$ acquires a real part (because in that case, the exponential $e^{\pm i \Re(\omega) t}$ oscillates in time), which occurs along the boundary defined by:
\begin{align}
\tilde A = \frac{1}{\tilde K^o} - \tilde K^o. \label{exceptional}
\end{align}
 
The corresponding eigenvectors are: 
\begin{align}
    u_i &= \frac{\qty(1 \pm \sqrt{1 - \tilde K^o \tilde A -(\tilde K^o)^2})\hat q_i  - (\tilde A+ \tilde K^o) \epsilon_{ji}\hat q_j}{\sqrt{2 \qty(1 \pm \sqrt {1 -\tilde K^o \tilde A - (\tilde K^o)^2})+\tilde K^o \tilde A + \tilde A^2 }}. \label{eigvecs} 
\end{align}
From Eq. (\ref{eigvecs}), we see that the two branches have eigenvectors which are no longer orthogonal. Indeed, for non-vanishing activity, the dynamical matrix in Eq.~(\ref{matrix}) is non-Hermitian due to the injection of energy by active components.  
Note that Eq.~(\ref{eigvecs}) reveals an important feature about the transition towards the propagation of elastic waves, described by Eq.~(\ref{exceptional}). This transition is characterized by two coinciding features: (i) the spectrum acquires a degeneracy and (ii) the corresponding eigenvectors become co-linear. Such transitions are known as exceptional points~\cite{Bender1998,Heiss2012}. Above the exceptional point, perpendicular and parallel components of the eigenvector become out of phase, and thus the displacement field has circular motion (see S.I. movies). 

An insightful exercise is to calculate the spectrum in the limit $\abs{K^o} \gg B, \mu, A$. In this case, the displacement vector $u_i$ traces out a circle as a function of time. Suppose that this circle has radius $R$. Then for a wave with wave vector $q_i$, the strain takes the form $u_{ij} = i q_{i} u_j$. Hence, a circle of radius $q R$ will be traced out in $S_1$-$S_2$ space, so the work done by the odd elastic material in a single cycle is $2\pi K^o q^2 R^2 $. The energy dissipated due to viscosity on a single cycle of period $T = \frac{2 \pi}\omega$ is given by $\eta \abs{\dot u}^2 T = 2 \pi \omega R^2$. Hence balancing the energy injected with the energy dissipated gives $\omega = \frac{K^o}{\eta} q^2$, which agrees with Eq.~(\ref{Spec}).

\subsection{Microscopic spectrum} In the overdamped regime, the linearized microscopic equation of motion is:
\begin{align}
    -i \gamma \omega u_i (\vb q) = D_{ij} (\vb q) \tilde u_j (\vb q),
\end{align}
where $D_{ij} (\vb q)$ is the dynamical matrix and $\gamma$ is a microscopic drag coefficient that is related to the macroscopic drag coefficient by $\eta = \frac{\gamma}V$, where $V$ is the area of the unit cell. 

For an unbounded triangular lattice with interactions of the form in Eq.~(\ref{Force}), we obtain an analytic form for the spectrum:
\begin{widetext}
\begin{align*}
\omega(q_x,q_y) = &- i \bigg (3 -\cos(q_x) - 2 \cos(\frac{q_x}2) \cos(\frac{\sqrt3 q_y}2) 
\pm \frac1{\sqrt2} \bigg\{ 3 - \cos(q_x) + \cos(2 q_x) - 2 \cos(\frac{q_x}2) \cos(\frac{\sqrt3 q_y}2)  \\
&- 
 2 \cos(\frac{3 q_x}2) \cos(\frac{\sqrt3 q_y}2)-\cos(\sqrt3 q_y) + 2 \cos(q_x) \cos(\sqrt3 q_y)
 + 3r^2 \bigg[ -6 + 3 \cos(q_x) \\ & + 6 \cos(\frac{q_x}2) \cos(\frac{\sqrt3 q_y}2)
 -2 \cos(\frac{3 q_x}2) \cos(\frac{\sqrt3 q_y}2) - \cos(\sqrt3 q_y)\bigg] \bigg\}^{1/2} \bigg ), 
\end{align*}
where for convenience we set the lattice spacing, spring constant $k$, and drag coefficient $\eta$ all equal to 1 and we define $r \equiv k^o/k$. 
\end{widetext}

\pagebreak

\bibliographystyle{naturemag}
\bibliography{oddelasticity}

\end{document}